\documentclass[twocolumn]{aastex6}
\usepackage{times}
\usepackage{amsmath}
\usepackage{textcomp}
\usepackage{amssymb}
\usepackage{natbib}
\usepackage{float}
\usepackage{color}
\usepackage{graphicx}
\usepackage{epstopdf}
\newcommand{\fermi}{{\it Fermi}}

\newcommand{\phflux}{\mbox{${\rm \, ph \,\, cm^{-2} \, s^{-1}}$}}

\newcommand{\gm}{$\gamma$}
\slugcomment{ApJ Letters accepted}

\shorttitle{Stacking Analysis of Extreme Blazars}
\shortauthors{Paliya et al.}

\begin{document}
\title{{\it Fermi}-LAT Stacking Analysis Technique: An Application to Extreme Blazars and Prospects for their CTA Detection}

\author{Vaidehi S. Paliya$^{1}$, A. Dom{\'{\i}}nguez$^2$, M. Ajello$^{3}$, A. Franckowiak$^1$, and D. Hartmann$^{3}$} 
\affil{$^1$Deutsches Elektronen Synchrotron DESY, Platanenallee 6, 15738 Zeuthen, Germany}
\affil{$^2$IPARCOS and Department of EMFTEL, Universidad Complutense de Madrid, E-28040 Madrid, Spain}
\affil{$^3$Department of Physics and Astronomy, Clemson University, Kinard Lab of Physics, Clemson, SC 29634-0978, USA}

\email{vaidehi.s.paliya@gmail.com}

\begin{abstract}
We present a likelihood profile stacking technique based on the \fermi-Large Area Telescope (LAT) data to explore the \gm-ray characteristics of \fermi-LAT undetected astrophysical populations. 
The pipeline is applied to a sample of \gm-ray unresolved extreme blazars, i.e., sources with the highest synchrotron peak frequencies ($\nu_{\rm Syn}^{\rm peak}\geqslant 10^{17}$ Hz), and we report a cumulative \gm-ray detection with more than 32$\sigma$ confidence for 2 degrees of freedom. Comparing the generated stacked \gm-ray spectrum with the sensitivity limits of the upcoming Cherenkov Telescope Array (CTA), we find that the \fermi-LAT undetected population of such extreme blazars, on average, may remain well below the CTA detection threshold due to their faintness and extragalactic background light (EBL) absorption. However, \gm-ray detected blazars belonging to the same class are promising candidates for CTA observations. The EBL corrected stacked spectra of these sources do not show any softening up to 1 TeV. This finding suggests the inverse Compton peak of extreme blazars to lie above 1 TeV, thus indicating a hard intrinsic TeV spectrum. Our analysis also predicts that at 100 GeV, at least $\sim$10\% of the diffuse extragalactic \gm-ray background originates from the \gm-ray undetected extreme blazars. These results highlight the effectiveness of the developed stacking technique to explore the uncharted territory of \gm-ray undetected astrophysical objects.

\end{abstract}

\keywords{methods: data analysis --- gamma rays: general --- galaxies: active --- galaxies: jets --- BL Lacertae objects: general}

\section{Introduction}{\label{sec:Intro}}
The Large Area Telescope (LAT) onboard the {\it Fermi Gamma ray Space Telescope} has revealed various types of astrophysical objects as \gm-ray emitters. The most numerous of them are blazars, i.e., radio-loud quasars with powerful relativistic jets pointed towards the observer, followed by narrow-line Seyfert 1 galaxies, pulsars, and many others \citep[cf.][]{2019arXiv190510771T}. Concerning blazars, the \fermi-LAT observations have allowed us to explore various unsolved problems related to jet physics and/or \gm-ray astronomy in general. A few examples are: the cosmic evolution of blazar jets \citep[][]{2012ApJ...751..108A,2014ApJ...780...73A}, connection of the central engine (i.e., central black hole and the accretion disk) with the relativistic jet \citep[e.g.,][]{2016ApJ...825L..11P,2017ApJ...851...33P}, the measurement of the Extragalactic Background Light \citep[EBL,][]{2015ApJ...813L..34D,2018Sci...362.1031F}, and the contribution of \gm-ray blazars to the Extragalactic Gamma-ray Background \citep[EGB,][]{2015ApJ...799...86A,2015ApJ...800L..27A}.

There are probably astrophysical source populations that are yet to be detected in the \gm-ray band. They are, similar to the \fermi-LAT detected objects, crucial to study their contribution to the diffuse \gm-ray background emission. Focusing on blazars, a characterization of the \gm-ray properties is also pivotal to determine their detectability with the next generation high-energy missions, e.g., the All-sky Medium Energy Gamma-ray Observatory \citep[AMEGO,][]{2017ICRC...35..798M} and the Cherenkov Telescope Array \citep[CTA,][]{2011ExA....32..193A}. In particular, AMEGO (with an energy range 200 keV to $\gtrsim$10 GeV) is expected to discover some of the most powerful blazars, especially at high-redshifts \citep[$z>3$,][]{2019arXiv190306106P}. On the other hand, CTA, which will operate in the $\sim$0.02$-$300 TeV band, will observe the most efficient particle accelerator jets from BL Lac objects, a sub-class of blazars with no or weak optical emission lines \citep[][]{1991ApJ...374..431S}, along with other types of sources, e.g. flat-spectrum radio quasars. Until then, the high-energy properties of these peculiar objects can be explored using \fermi-LAT observations.

A useful methodology to explore the characteristics of any astrophysical populations, especially undetected ones, is the stacking technique. This has been successfully applied earlier to Energetic Gamma-Ray Experiment Telescope data to search for the \gm-ray signal from, e.g., cluster of galaxies \citep[][]{2003ApJ...588..155R} and infrared (IR) galaxies \citep[][]{2005ApJ...621..139C}. Also in the \fermi-LAT era, various stacking algorithms have been developed to search for \gm-ray emission from undetected populations. \citet[][]{2012A&A...547A.102H} proposed a method which co-adds the \fermi-LAT count maps and performs a maximum likelihood analysis on the combined data to derive the \gm-ray parameters. The \fermi-LAT data analysis software, {\it Fermitools} provides a package, {\tt Composite2}, which makes use of summed log-likelihood functions to combine the likelihood fitting of multiple sources at once. This tool ties together the spectral parameters of interest for all sources under consideration before performing the fit and
has been used in some studies \citep[e.g.,][]{2011PhRvL.107x1302A,2014PhRvD..89d2001A}.

Here we present a new approach based on the stacking of the individual source likelihood profiles to explore the \gm-ray properties of the \fermi-LAT undetected population. The developed technique is sensitive to extract the faint \gm-ray signal, is quick, and is flexible to be used for any kind (binned or unbinned) of \fermi-LAT data analysis. It can also be applied to \gm-ray detected sources to estimate the average properties of the population. 
Unlike {\tt Composite 2}, the tool can be used to independently generate the likelihood profiles of as many sources as one needs before combining them, at the expense of parallel processing computational resources. We describe the steps of the stacking technique in Section~\ref{sec2} and present the results of its application to a sample of extreme blazars in Section~\ref{sec3}. The results associated with the validation of the stacking technique are presented in Section~\ref{sec4}.  In Section~\ref{sec5}, we discuss and summarize our findings. All the quoted uncertainties are estimated at 1$\sigma$ level, unless specified.

\section{The Pipeline}\label{sec2}
The input for the stacking pipeline is the list of sky positions, i.e., right ascension and declination, for the objects under consideration. The tool works in the following two steps:

\subsection{Pre-processing}
First, a standard likelihood analysis is performed on the \fermi-LAT data covering a given time period, energy range, and a selected region of interest (ROI) using {\tt fermiPy} \citep[][]{2017arXiv170709551W}. 
%The data are binned with a pixel size of 0$^{\circ}$.05 and 10 bins per energy decade. 
To model the \gm-ray sky, we consider the recently released fourth catalog of \fermi-LAT detected sources \citep[4FGL,][]{2019arXiv190210045T}.
% which presents the list of the \gm-ray emitters detected in the first 8 years of \fermi~operation in the energy range of 50 MeV to 1 TeV. 
 We use the latest interstellar emission model and the standard template for the isotropic emission\footnote{https://fermi.gsfc.nasa.gov/ssc/data/access/lat/BackgroundModels.html}. All 4FGL sources lying within ROI size+$R$ are considered in the likelihood fitting, where size of the ROI and additional radius $R$ depend on the minimum energy chosen for the analysis. Spectral parameters associated with the sources lying within the ROI are allowed to vary during the likelihood fitting and are kept fixed to the 4FGL values if they lie outside the ROI. After a first round of the optimization, we scan the ROI to search for unmodeled \gm-ray objects by generating test statistic (TS) maps. The maximum likelihood TS is defined as TS =  $2\log(\mathcal{L}_1-\mathcal{L}_0$), where $\mathcal{L}_0$ and $\mathcal{L}_1$ denote the likelihood values without (i.e., null hypothesis) and with (alternative hypothesis) a point source at the position of interest, respectively \citep[][]{1996ApJ...461..396M}. Since unmodeled sources can have hard or soft spectra, TS maps are generated for various photon indices, e.g., 1.5, 2, 2.5. When an excess emission (TS$>$25) is found, it is added to the sky model following a power law spectral model and a second set of TS maps is generated. Once all excesses above the background are identified and inserted to the model, a final likelihood fit is performed to optimize all the free parameters in the ROI and derive the spectral parameters for the source of interest. This exercise enables us to segregate the whole sample in the \gm-ray detected and undetected objects.

\begin{figure*}[t!]
\hbox{
\includegraphics[scale=0.425]{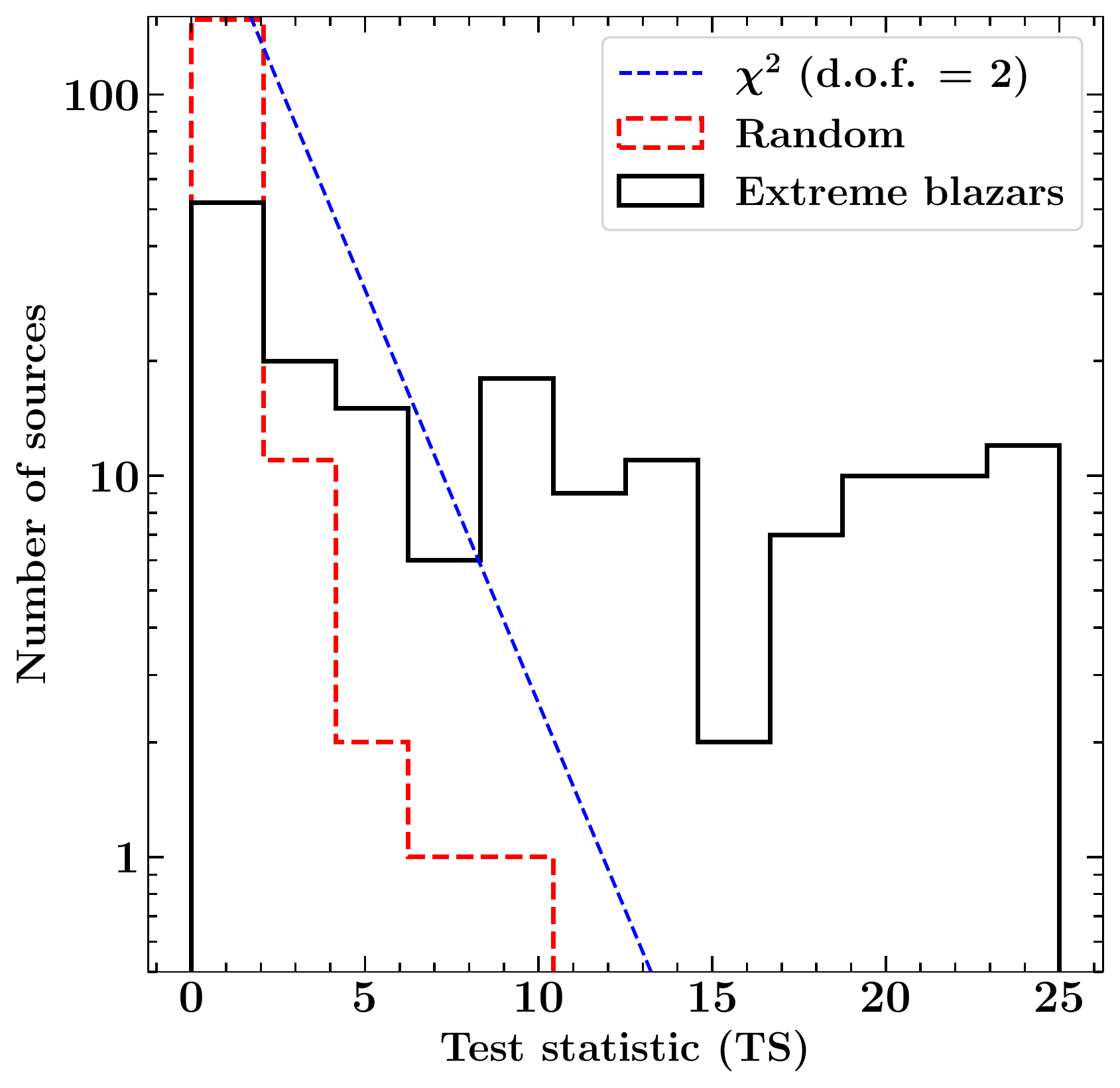}
\hspace{0.5cm}
\includegraphics[scale=0.45]{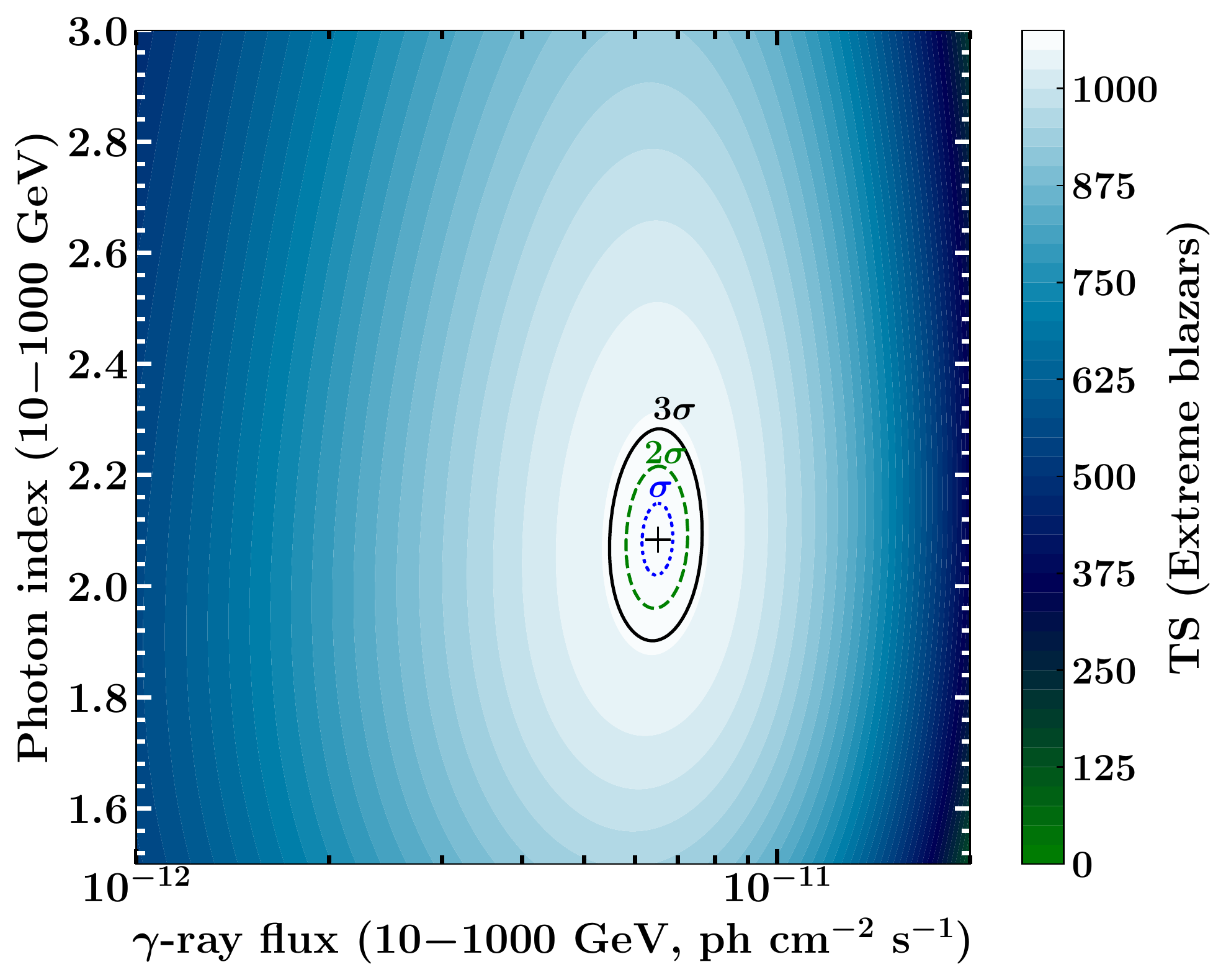}
}
\caption{Left: The TS histograms of the 172 extreme BL Lac objects (black solid) and empty sky positions representing the background (red dashed). We also plot the $\chi^2$ distribution for 2 degrees of freedom (blue dashed line) as a representation of the null hypothesis. Right: Stacking analysis of extreme blazars. Confidence contours at $\sigma$, 2$\sigma$, and 3$\sigma$ levels are shown.} \label{fig:stacking}
\end{figure*}

\subsection{Stacking}
The pipeline considers the source lists generated in the previous step and proceeds as follows:
\begin{enumerate}

\item Assuming that the average spectral behavior of the source population is well represented with a power law model, we generate a grid of photon flux ($10^{-15}-10^{-8}$ \phflux~in 50 logarithmic steps) and photon index (1.5$-$3.5 in the interval of 0.1). Note that the lower limit of the photon flux grid should be small enough to represent the absence of any \gm-ray source so that the likelihood value computed at this photon flux refers to the null hypothesis of the \gm-ray detection. 

\item The photon flux and index at a given grid point are then used as the source model parameters and the fitting is performed to determine the log-likelihood value. This step is repeated at every grid point, thus effectively generating a likelihood profile. To speed-up the process, we freeze the spectral parameters of all other modeled sources to the optimized values derived in the previous step, except the diffuse background models which are allowed to vary. By subtracting the log-likelihood value at the lowest flux (representing the null hypothesis) from the generated profile, we compute the TS or detection significance profile for a given source. This step is repeated for all \gm-ray undetected sources in the sample and a set of TS profiles is created.

\item Since the log-likelihood is additive in nature, we add the generated TS profiles to produce a combined profile representing the whole sample. The TS peak position and 1$\sigma$ uncertainties in the associated spectral parameters are then estimated by fitting a spline function. The photon flux and index values associated with the TS peak represent the average spectral parameters of the whole population. 
\end{enumerate}

 \begin{figure*}[t!]
\hbox{
\includegraphics[scale=0.45]{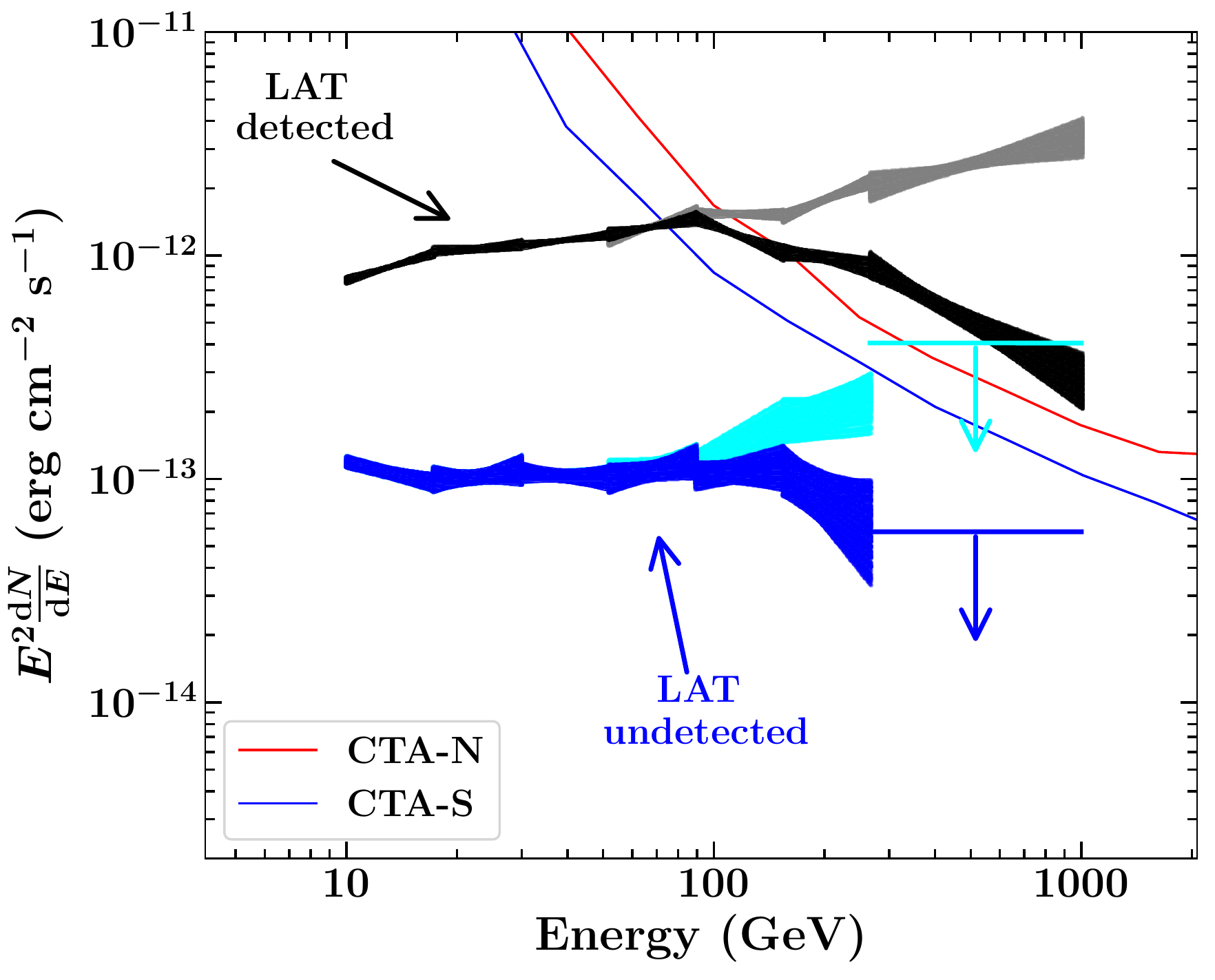}
\includegraphics[scale=0.48]{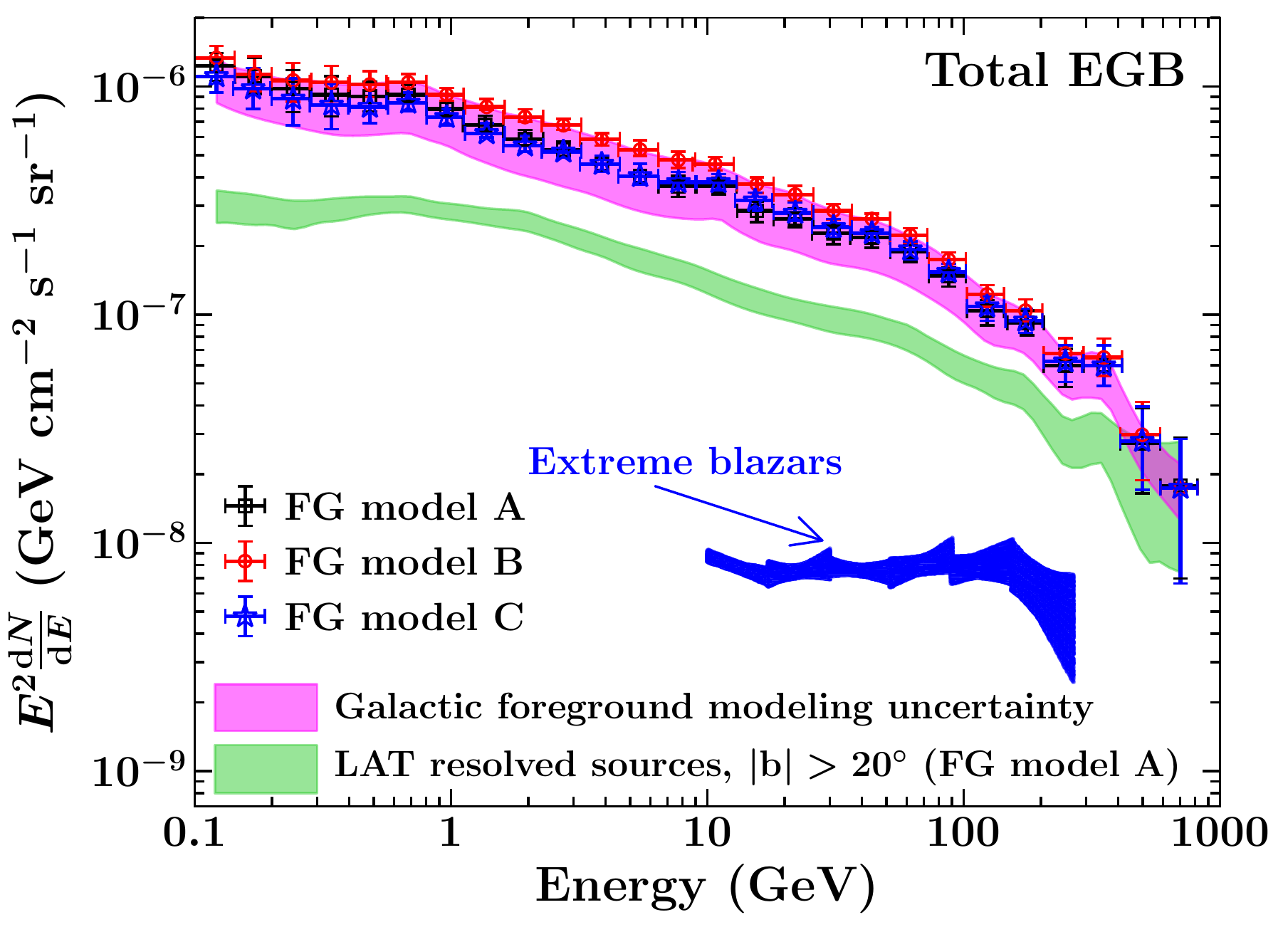}
}
\caption{Left: Gamma-ray stacked SED of extreme blazars as labeled. In each energy bin, we show the spectrum with a bow-tie plot if the peak TS is $>$25. The shown upper limit is estimated at 95\% confidence level. Note the faintness of the undetected population with respect to CTA sensitivity limits. Right: Comparison of the cumulative EGB intensity from the \gm-ray undetected extreme blazars with the total EGB. Black, red, and blue data points refer to the EGB intensities measured for different Galactic foreground (FG) models discussed in \citet[][]{2015ApJ...799...86A}. The EGB contribution by the high-latitude \fermi-LAT resolved sources is shown with green shaded area  \citep[see][for details]{2015ApJ...799...86A}.} \label{fig:sed}
\end{figure*}

\section{Application to Extreme Blazars}\label{sec3}
We apply the developed stacking technique to a sample of extreme blazars and use them to demonstrate the robustness of the pipeline by performing simulations and background checks.

Extreme blazars are a sub-sample of BL Lac objects that have the synchrotron peak\footnote{The spectral energy distribution (SED) of a blazar is characterized by a double hump structure. The low-energy peak, usually located in the sub-mm-to-X-ray bands, originates from synchrotron emission, whereas, the high-energy peak is usually explained by inverse Compton scattering off low energy photons by the relativistic electrons present in the jet \citep[see, e.g.,][for a review]{2019Galax...7...20B}.} located at very high frequencies ($\nu_{\rm Syn}^{\rm peak}\geqslant 10^{17}$ Hz) indicating them to host some of the most efficient particle accelerator jets \citep[e.g.,][]{2001A&A...371..512C}. A high synchrotron peak frequency also suggests their inverse Compton peak to be located at very high energies (VHE, $>$100 GeV) making them bright TeV candidates and a promising source population for CTA detection. However, the same phenomenon causes the extreme blazars to be faint and less variable in the \fermi-LAT energy range \citep[e.g.,][]{2011MNRAS.414.3566T,2018MNRAS.480.2165A}. Therefore, it is instructive to perform a stacking analysis of the \gm-ray undetected extreme blazars and determine whether they are GeV \gm-ray emitters as a whole. The derived spectral parameters can be used to explore the detectability of extreme blazars with CTA and also use them to probe the EBL and the extragalactic VHE background \citep[][]{1993MNRAS.260L..21P,2015ApJ...800L..27A}.

We select 337 extreme blazars with known redshift from a sample of 2011 high-synchrotron peaked \citep[HSP, $\nu_{\rm Syn}^{\rm peak}\geqslant 10^{15}$ Hz,][]{2010ApJ...716...30A} BL Lac objects included in the third catalog of HSP blazars (3HSP, \citealt{Chang19}, see also \citealt{2017A&A...598A..17C})\footnote{http://www.ssdc.asi.it/3hsp/}. The sample has a redshift range of 0.01$-$0.85 with a mean redshift of 0.35. We use the P8R3 LAT data acquired in $\sim$127 months (2008 August 4 to 2019 March 18) of \fermi~operation and in the energy range of 10$-$1000 GeV for the analysis. The minimum energy is chosen as 10 GeV to allow a maximum overlap with the frequency band covered by CTA. Furthermore, the \fermi-LAT PSF considerably improves above 1 GeV \citep[e.g.,][]{2013arXiv1303.3514A} thereby enabling a better source localization and suppressing of the diffuse background emission which is bright at MeV energies \citep[cf.][]{2016ApJS..223...26A}. We use a squared ROI of $10^{\circ}\times10^{\circ}$ and adopt a zenith angle cut ($z_{\rm max}$) of 105$^{\circ}$. With these settings, the {\it Pre-processing} led to a significant \gm-ray detection of 165 extreme blazars\footnote{Among 165 \gm-ray detected objects, 154 are also present in the 4FGL catalog.}. The TS distribution of the remaining 172 \gm-ray undetected extreme blazars is shown in the left panel of Figure~\ref{fig:stacking}. This is consistent with the null hypothesis or background fluctuations, i.e., $\chi^2$ distribution with 2 degrees of freedom, at low TS and shows excess above TS$\gtrsim$10 that can be attributed to the jet emission.

 \begin{figure*}[t!]
\hbox{
\includegraphics[scale=0.355]{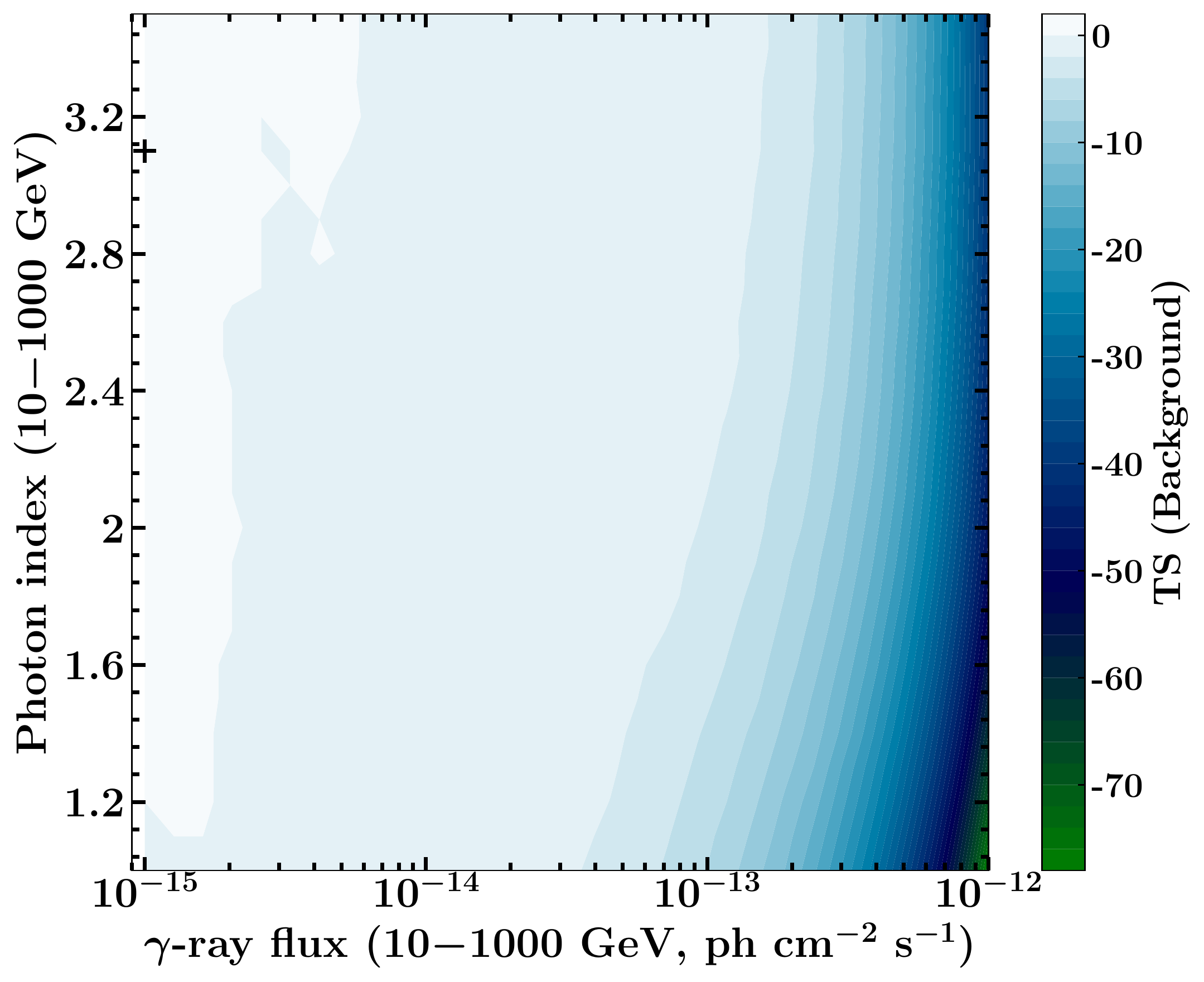}
\hspace{0.5cm}
\includegraphics[scale=0.4]{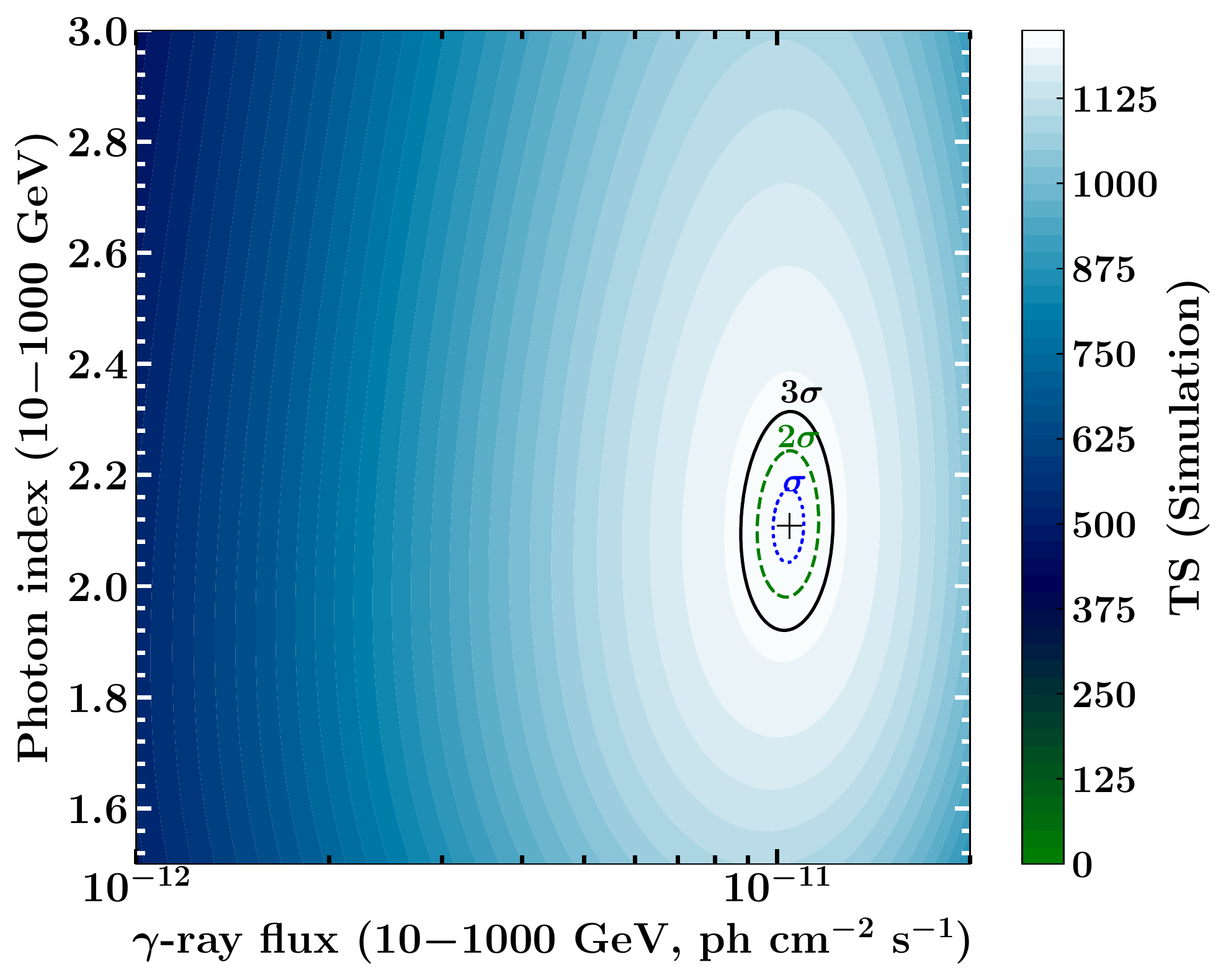}
}
\caption{Left: Stacking analysis of randomly selected 172 empty sky positions. The negative TS observed at large \gm-ray fluxes implies that the alternative hypothesis of the presence of a \gm-ray source with the given photon flux and index is strongly rejected with respect to the null hypothesis of no detection. Right: Same as left, but for 100 simulated \gm-ray sources with \gm-ray spectral properties similar to real extreme blazars.} \label{fig:test}
\end{figure*}

The stacking pipeline is then applied to the 172 extreme blazars and we show the results in the right panel of Figure~\ref{fig:stacking}. As can be seen, the stacked emission is well detected by the \fermi-LAT with a TS = 1062 ($\sim32\sigma$ significance for 2 degrees of freedom). The associated peak \gm-ray flux and photon index are $F_{\rm 10-1000~GeV}=6.51^{+0.36}_{-0.35}\times 10^{-12}$ \phflux~and $\Gamma_{\rm 10-1000~GeV}=2.08^{+0.07}_{-0.06}$. These results demonstrate the capabilities of the stacking technique to extract the faint \gm-ray signal from a \fermi-LAT undetected population.

The strong \gm-ray signal has also allowed us to perform the stacking in smaller energy bins and generate a stacked \gm-ray spectrum. Using the best-fit photon flux and index derived in each of the seven energy bins, we make the bow-tie plot and show the cumulative spectrum in the left panel of Figure~\ref{fig:sed}. In this diagram, we also show the stacked \gm-ray spectrum of 165 \fermi-LAT detected extreme blazars generated using the same pipeline. For a comparison, we overplot sensitivity limits of CTA for 50 hr integration time\footnote{https://www.cta-observatory.org}. 

The \gm-ray spectrum of undetected sources remains hard up to $\sim$100 GeV ($\frac{{\rm d}N}{{\rm d}E}\propto E^{-2}$) and declines after, most likely due to EBL absorption. The average spectrum of the \fermi-LAT detected sources also reveals a well defined peak at $\sim$100 GeV and steepens at higher energies. To explore the role of the EBL, we generate stacked spectra of both populations taking into account the EBL absorption \citep[][]{2011MNRAS.410.2556D}. The results are plotted in Figure~\ref{fig:sed} (left panel) with cyan and gray colors for \gm-ray undetected and detected extreme blazars, respectively. Since the EBL corrected stacked spectra do not exhibit any softening, the inverse Compton peak of the \fermi-LAT detected population lies above 1 TeV, thus indicating a very hard intrinsic TeV spectrum\footnote{We have also tested whether the shape of the stacked SED is dominated by a few bright sources. This is done by normalizing the SED of each blazar by its integrated 10$-$1000 GeV flux. We find that such a normalized stacked spectrum has a shape similar to that plotted in Figure~\ref{fig:sed} and therefore the results shown here truly represent the average behavior of the population.}. A similar result holds for the \gm-ray undetected sources though a strong claim cannot be made due to the flux upper limit in the highest energy bin.

The derived results also demonstrate the effectiveness of the developed stacking technique to extract the \gm-ray signal from a population which is  $\sim$an order of magnitude fainter than the \fermi-LAT detected one. By comparing this signal with the plotted sensitivity limits, we find that the probability of CTA detection is higher for objects that already have \fermi-LAT detections (Figure~\ref{fig:sed}). The \gm-ray undetected extreme blazar population, on the other hand, remains below the detection limit of CTA both due to their low level of \gm-ray emission and the EBL absorption \citep[see also][]{2019arXiv190703666F}. This result does not rule out possibility of CTA detection for a few individual sources, especially during a TeV flaring state. However, considering the source population as a whole, we show that \fermi-LAT detected extreme blazars are better candidates for VHE observations.

With the knowledge of the average \gm-ray spectrum, we can derive the contribution of the \fermi-LAT undetected extreme blazars to the total EGB. It is computed by assuming that the sources are uniformly distributed in the sky outside the Galactic plane\footnote{All extreme blazars studied in this work lie outside the Galactic plane. The assumption of the uniform sky distribution allows us to derive the lower limit to the EGB contribution by the astrophysical population under consideration. The accurate measurement can be done by considering the source count distribution or luminosity function \citep[see, e.g.,][]{2016PhRvL.116o1105A} and requires a precise estimation of the solid angle covered in the sky. These aspects, however, are beyond the scope of this work.} ($\vert{\rm b}\vert>10^{\circ}$). In the right panel of Figure~\ref{fig:sed}, we show the derived results and compare them with the total EGB estimated in \citet[][]{2015ApJ...799...86A}. According to our analysis, \gm-ray undetected extreme blazars are responsible for at least $\sim$10\% of the total EGB at 100 GeV.

\section{Validation of the Stacking Technique}\label{sec4}
\subsection{Background Stacking}
Since we combine the \gm-ray undetected objects, a genuine question arises about the possibility of stacking the diffuse background emission rather than the radiation originated from actual point sources. Therefore, to test the robustness of the \gm-ray signal reported above, we randomly select 172 empty sky positions not lying within 95\% error radius of any 4FGL source and repeat the same procedure as adopted for extreme blazars. In the left panel of Figure~\ref{fig:stacking}, we show the TS histogram of these random positions. Comparing the distribution with the null hypothesis ($\chi^2$ curve), it can be concluded that the derived signal is fully compatible with the random background fluctuations. The corresponding stacking plot shown in the left panel of Figure~\ref{fig:test} confirms the negligible contribution of the random background fluctuations to the stacked \gm-ray emission observed from extreme blazars. 

\subsection{Stacking of Simulated Sources}
The robustness of the stacking technique is also verified by performing simulations of $\sim$127 months of the \fermi-LAT data for 100 objects. The \gm-ray spectra of these sources are assumed to follow a power law. The power law indices are randomly extracted from a Gaussian distribution peaking at 2.1 with a dispersion of 0.1. On the other hand, fluxes are selected from a log-normal distribution having the peak at 10$^{-11}$ \phflux~and a dispersion of 0.1 in log-space. We populate the background sky with 4FGL sources and Galactic and isotropic diffuse emissions and repeat the entire procedure as described above. The results of the stacking of the simulated objects are presented in the right panel of Figure~\ref{fig:test}. We find the best-fit photon flux and index values as $1.04^{+0.01}_{-0.01}\times10^{-11}$ \phflux~and 2.11$^{+0.07}_{-0.07}$, respectively at peak TS = 1216. Clearly, the agreement between the input and output spectral parameters confirms the effectiveness of the developed tool in measuring the average \gm-ray properties of the unresolved population.

\section{Discussion and Summary}\label{sec5}
We have described a stacking technique to extract the \gm-ray signal from the \fermi-LAT undetected sources by combining their likelihood profiles. The pipeline is applied to a sample of 172 extreme blazars resulting in a strong \gm-ray detection of the population. The tool is capable of extracting about an order of magnitude fainter \gm-ray signal than that possible with the conventional point source \fermi-LAT data analysis. A crucial finding of the stacking analysis is that CTA may not be able to detect VHE emission from these objects mainly due to their faintness and EBL absorption. However, sources that already have been detected by \fermi-LAT should be primary targets for CTA observations. Another important finding is that the EBL-corrected stacked spectrum of extreme blazars exhibits no softening up to 1 TeV, thus indicating their inverse Compton peak to lie at $>$1 TeV. At 100 GeV, a significant fraction ($\sim$10\%) of the total EGB originate from the \gm-ray undetected extreme blazars.

The simplicity of the developed technique offers a possibility to apply it to various astrophysical problems other than just to determine the \gm-ray detection/non-detection of a source population. One such example could be the known correlation between the IR and \gm-ray luminosities ($L_{\rm IR}$ and $L_{\gamma}$, respectively) in star-forming galaxies \citep[e.g.,][]{2012ApJ...755..164A}. Since $L_{\rm IR}$ is well-known, the stacking technique can be used to explore the strength of the correlation and associated parameters that maximizes the TS profile for a given set of \gm-ray spectral parameters (and thus $L_{\gamma}$). This is fully explored in a companion paper where we apply the developed pipeline to a sample of star-forming galaxies (Ajello et al., in preparation). Finally, the stacking technique is flexible to adopt any spectral models (other than power law used here), including EBL attenuated ones, and can be applied to any astrophysical populations, e.g., galaxy clusters and X-ray binaries \citep[cf.][]{2013A&ARv..21...64D}. This makes it a versatile tool for \gm-ray astronomy.

\acknowledgments
We are thankful to the referee for a constructive criticism. The \fermi-LAT Collaboration acknowledges support for LAT development, operation and data analysis from NASA and DOE (United States), CEA/Irfu and IN2P3/CNRS (France), ASI and INFN (Italy), MEXT, KEK, and JAXA (Japan), and the KA Wallenberg Foundation, the Swedish Research Council and the National Space Board (Sweden). Science analysis support in the operations phase from INAF (Italy) and CNES (France) is also gratefully acknowledged. This work performed in part under DOE Contract DE-AC02-76SF00515.

\software{fermiPy \citep{2017arXiv170709551W}}.

\bibliographystyle{aasjournal}
%\bibliography{Master}

\end{document}